\documentclass[prb,twocolumn,longbibliography]{revtex4-1}

\usepackage{graphicx}  
\usepackage{amsmath} 
\usepackage{amsfonts} 
\usepackage{hyperref}

\def\E{\mbox{e}}
\def\dd{\mbox{d}}

\def\dt{\mbox{d}t}

\def\d3r{\mbox{d}^3r}

\def\e0{\varepsilon_0}

\def\beq{\begin{equation}}
\def\eeq{\end{equation}}
\def\bea{\begin{eqnarray}}
\def\eea{\end{eqnarray}}
\def\bi{\begin{itemize}}
\def\ei{\end{itemize}}
\def\bp{\begin{picture}}
\def\ep{\end{picture}}

\def\betav{\mbox{\boldmath$\beta$}}

\def\nablav{\mbox{\boldmath$\nabla$}}

\def\grad{\nablav}
\def\rot{\nablav\times}
\def\div{\nablav\cdot}

\def\ddt{\partial_t}
\def\ddx{\partial_x}

\def\ddtau{\partial_{\tau}}

\def\dd{\mbox{$\mathrm d$}}

\def\xv{\hat{\bf x}}

\begin{document}

\title{Plasma Waves in a Different Frame:\\ a Tutorial for Plasma-based Electron Accelerators
}

\author{A. Macchi}\email{andrea.macchi@ino.cnr.it}
\affiliation{National Institute of Optics, National Research Council (CNR/INO), Adriano Gozzini laboratory, Pisa, Italy}
\affiliation{Enrico Fermi Department of Physics, University of Pisa, Pisa, Italy}

\date{\today}

\begin{abstract}
Nonlinear, relativistic longitudinal waves with sub-luminal phase velocity $v_p$ are the basis of plasma-based electron accelerators. For such application, key properties of the wave are the maximum or ``wave breaking'' amplitude and the corresponding energy gained by electrons trapped in the wave field. Here we show that these properties and the general waveform are obtained with little mathematical complexity by using a Lorentz transformation to a frame co-moving at velocity $v_p$. The transformation reduces the problem to a second-order ordinary differential equation as originally found by Chian [Plasma Phys. \textbf{21}, 509 (1979)] so that the analysis can exploit the analogy with the mechanical motion of a particle in a potential well. This approach seems particularly suitable for a compact, tutorial introduction to plasma-based electron accelerators with little mathematical complexity. 
\end{abstract}

\maketitle

\section{Introduction}

The propagation of harmonic electromagnetic waves in a continuous medium can be described by the equation for the electric field ${\bf E}=\mbox{Re}\left[\tilde{\bf E}({\bf r})\mbox{e}^{-i\omega t}\right]$,
\bea
\nabla^2\tilde{\bf E}+\frac{\omega^2}{c^2}\varepsilon(\omega)\tilde{\bf E}-\grad(\div\tilde{\bf E})=0 \; ,
\label{eq:wave}
\eea
where the dielectric function $\varepsilon(\omega)$ describes the linear response of of the medium at the frequency $\omega$. For transverse plane waves satisfying $\div{\bf E}=0$, such that $\tilde{\bf E}={\bf E}_0\mbox{e}^{i{\bf k}\cdot{\bf r}}$ with ${\bf k}\cdot{\bf E}=0$, Eq.(\ref{eq:wave}) immediately provides the well-known dispersion relation $\omega^2=k^2c^2/\varepsilon(\omega)$ between $\omega$ and the wavevector $k$. If $\varepsilon(\omega)$ is a real and positive quantity, the wave propagates in the medium with phase velocity $v_p=\omega/k=c/{\sf n}(\omega)$ where ${\sf n}(\omega)=\varepsilon^{1/2}(\omega)$ is the refractive index.

A medium can also support electrostatic ($\rot{\bf E}=0$), \emph{longitudinal} plane waves having the electric field parallel to the wavevector. Since for such waves  $\grad(\div{\bf E})=\nabla^2{\bf E}$, Eq.(\ref{eq:wave}) immediately shows that $\varepsilon(\omega)=0$, i.e. their frequency is a zero of the dielectric function. The simplest example is that of a collisionless, ideal cold plasma or zero-temperature electron gas having an electron density $n_e$ with unperturbed (background) value $n_0$, for which $\varepsilon(\omega)=1-\omega_p^2/\omega^2$, where $\omega_p=(4\pi e^2n_0/m_e)^{1/2}$ is the plasma frequency. Such plasma wave (also refereed to as a bulk plasmon in solid state physics) is characterized by an oscillation of the electron density $n_e$, which is related to the electric field by the equation
\bea
\div{\bf E}=4\pi\rho=4\pi e(n_0-n_e) \; .
\label{eq:poisson}
\eea
Although longitudinal waves or plasma waves are straightforward solutions of Eq.(\ref{eq:wave}), apparently they are not presented in most textbooks of introductory electrodynamics and were only rarely and partially discussed in this journal\cite{germinAJP62,mirmanAJP64,arbabAJP65,chenAJP69,giovanielliAJP75,haleviAJP80,lippmannAJP88,boozerAJP10} so it is worth reminding some peculiar features.
For the plasma wave with $\omega=\omega_p$ the wavevector $k$ and the corresponding phase velocity $v_p=\omega_p/k$ are not determined by the dispersion relation, so that in principle one may excite a wave with any value of $v_p$. Moreover, the wave cannot be arbitrarily strong, because the amplitude of the density oscillation cannot exceed the background value, i.e. $|n_e-n_0| \leq n_0$. By writing the electric field of the plasma wave as ${\bf E}=\mbox{Re}\left[E_0\xv\mbox{e}^{ikx-i\omega_pt}\right]$ and using Eq.(\ref{eq:poisson}) along with the equation of motion $m_e\dd{\bf v}/\dt=-e{\bf E}$  one easily obtains that the limit on the density is equivalent to the amplitude of the oscillation velocity being smaller than the phase velocity, i.e. 
\bea
v_{\rm osc}=\frac{eE_0}{m_e\omega_p} \leq v_p \; ,
\label{eq:WB-NR}
\eea 
which is also equivalent to the peak field not exceeding an upper limit, $E_0 \leq m_e\omega_pv_p/e$.  
From the point of view of the fluid description of the electrons, the oscillation velocity cannot exceed the phase velocity because otherwise the trajectories of the fluid elements would intersect and produce a singularity in the density. An harmonic wave driven beyond the maximum amplitude limit will lose its regular, periodical form and is said to break.

The longitudinal character of the plasma wave and the possibility to determine the phase velocity by a proper excitation mechanism are the physical basis of plasma accelerators of electrons, first proposed by Tajima \& Dawson\cite{tajimaPRL79} in 1979 (for a popular description see Joshi\cite{joshiSA06} and references therein). Much like a surfer catching a sea wave, a test electron which at some instant has a velocity close to $v_p$ will receive a large acceleration by the plasma wave. If $v_p$ is close to, but does not exceed the speed of light $c$, the wave can accelerate relativistic electrons, with the additional advantage that such electrons may gain a large energy with a small change in their velocity, thus remaining in phase with the wave for a long time. Plasma waves with $v_p \lesssim c$ can be created in the wake of intense laser pulses or particle bunches propagated through the plasma. This approach to future linear accelerators has achieved several recent milestones\cite{gruenerP12}.
At the Stanford Linear Accelerator Center, accelerating gradients of tens of GV/m over a distance of $\lesssim$1~m have been demonstrated by using an electron bunch to drive the plasma wave\cite{blumenfeldN07,litosN14}. 
At CERN, acceleration in a plasma wave driven by a proton bunch has been shown\cite{adliN18}.
Very recently a new world record of acceleration up to $8$~GeV was obtained for the laser-based approach at the Lawrence Berkeley National Laboratory\cite{gonsalvesPRL19}. 

In their paper\cite{tajimaPRL79}, Tajima \& Dawson were interested in showing the potential of their proposal by estimating the maximum energy ${\cal E}_{\rm max}$ theoretically attainable for electrons. Such energy depends on the maximum field of the accelerating wave, for which Eq.(\ref{eq:WB-NR}) was used in Ref.\onlinecite{tajimaPRL79} to provide an upper limit. However, since a phase velocity $v_p$ very close to $c$ was considered, Eq.(\ref{eq:WB-NR}) does not provide a good estimate, because the oscillation velocity must be close to $c$ as well. This implies that the electron dynamics in the field of the wave is relativistic and thus highly nonlinear. As we discuss below, the maximum or ``wave-breaking'' field $E_{\rm WB}$ of a nonlinear, relativistic plasma wave is much higher than the non-relativistic value given by (\ref{eq:WB-NR}); when using the latter, the maximum electron energy is largely underestimated. Luckily for Tajima \& Dawson, their involuntarily pessimistic estimate for ${\cal E}_{\rm max}$ did not prevent their proposal to be very successful, although a relativistically correct evaluation would have been much more impressive!

The theory of nonlinear waves in plasma has a long history. 
A description of relativistic waves in cold plasmas, including the calculation of $E_{\rm WB}$ for the longitudinal wave was given by Akhiezer \& Polovin\cite{akhiezer-polovinJETP56} in 1956. 
An extremely detailed and extended theoretical description of various nonlinear waves was given by Decoster\cite{decosterPR78}, and a review focusing on relativistic effects was given by Shukla et al.\cite{shuklaPR86}. 
The proposal of plasma-based acceleration greatly stimulated this research area (see e.g. the overview by Esarey et al.\cite{esareyIEEE96}): as an example, the maximum energy ${\cal E}_{\rm max}$ for an electron trapped in the wave has been calculated by Esarey \& Pilloff\cite{esarey-pilloffPoP95} using an Hamiltonian approach. In most of these brilliant theoretical works the calculations are not particularly straightforward, so that they may represent a barrier for newcomers in the field and any physicist interested in the results but unwilling to follow the mathematical derivation. It seems before of interest to present a compact calculation which does not use high level mathematics and should be accessible even to undergraduate physics students having a basic background in electrodynamics, mechanics, and special relativity. In fact, the calculation becomes quite easy in a reference frame co--moving with the wave at the velocity $v_p$. In such co--moving frame, where the fields are time-independent,\footnote{It may be interesting to notice that for waves with super-luminal phase speed $v_p>c$, in a moving frame with $\beta=c/v_p$ the fields are space-independent. The Lorentz transformation to such frame can also be helpful to simplify calculations\cite{winklesPoF72,decosterPR78}},
the properties of the wave can be obtained via a Newton--like second order ordinary differential equation (ODE) which was previously found by Chian\cite{chianPP79}. The analysis thus exploits the similarity with the motion of a particle in a potential well.\footnote{In Chian's paper an immobile ion background is assumed and this is also the only case here considered. A generalization of Chian's equation to mobile ions is given by Decoster\cite{decosterPR78}.} 

It is interesting to notice that the co--moving frame should be familiar to the plasma acceleration community since it was used by Tajima \& Dawson\cite{tajimaPRL79} to estimate ${\cal E}_{\rm max}$ (see section~\ref{sec:linear} below), but Chian's result seems to have been overlooked in this context. Here we recover and extend Chian's original discussion to quickly obtain $E_{\rm WB}$ and ${\cal E}_{\rm max}$ as well as to infer the nonlinear profile of the longitudinal wave, which in general cannot be obtained analytically. This presentation may thus be useful for tutorials and primers in laser-plasma acceleration as well as to illustrate a specific case of nonlinear ODE analyzed by the pseudopotential approach.

\section{Linear case}
\label{sec:linear}

First we give a description of a linear, harmonic plasma wave in the co-moving frame. The only non-zero field component in the laboratory frame $S$ is (we use directly the fields in complex form for brevity)
\bea 
E_x=E_0 \E^{ikx-i\omega_p t} \; .
\eea
Assuming $v_p=\omega_p/k<c$ and performing a Lorentz boost transformation for the four-vector $(\omega,k)$ (we write four-vectors omitting the trivial $y$ and $z$ components) with $\betav=(v_p/c)\xv$ to the frame $S'$, the frequency and wavevector become 
\bea
\omega' &=&\gamma_p(\omega_p-kv_p)=0 \; , \label{eq:w'} \\
k'&=&\gamma_p\left(k-\frac{\omega_pv_p}{c^2}\right)=\frac{k}{\gamma_p} \; ,
\label{eq:k'}
\eea
where $\gamma_p=(1-\beta_p^2)^{-1/2}$. Eqs.(\ref{eq:w'}--\ref{eq:k'}) are
in agreement with the invariance of $\omega^2-k^2c^2$.  
Moreover, in $S'$ the electric field is has the same amplitude as in $S$, since the field components parallel to $\betav$ are unchanged by a Lorentz transformation. We thus obtain 
\bea
E'_x=E_0 \E^{ik'x'} \; .
\eea
The charge and current densities are
\bea
\rho'=\frac{ik'}{4\pi}E_0\E^{ik'x'} \; , \qquad J'_x=0  \; ,
\label{eq:rho-Jx}
\eea
as obtained either directly from the equations $4\pi\rho'=\ddx'E$ and $4\pi J'_x=-\ddt E'_x$ or by performing the  Lorentz transformation for the four-current $(c\rho,J_x)$. 

It is instructive to check that $J'_x=0$ (a condition that will be used below) by studying the electron dynamics in $S'$.  
The current density can be written as a function of the ion and electron densities ($n'_{(i,e)}$) and velocities ($u'_{(i,e)x}$),
\bea
J'_x &=& e(n'_iu'_{ix}-n'_eu'_{ex})=e(-n'_0v_p-n'_eu'_{ex}) \nonumber \\
     &=& e(-n_0\gamma_pv_p-n'_eu'_{ex})
\label{eq:current}
\eea
since in $S'$ the background ions have density $n_0\gamma_p$ and flow with velocity $-v_p$. By posing $J'_x=0$ we obtain that the electrons must have a steady velocity $u'_{ex}=-(n_0/n'_e)v_p$. Of course, this velocity must be consistent with the electron equation of motion. At this point we must remember that $u'_{ex}=u'_{ex}(x)$ is a velocity field and that even if $u_{ex}\ll c$ in $S$, adding the boost velocity $-v_p$ in $S'$ make the electrons relativistic. 
We thus write for the electron momentum $p'_{ex}=m_e\gamma(u'_{ex})u'_{ex}$ (with $\gamma(u)=(1-u^2/c^2)^{-1/2}$)
\bea
\frac{\dd}{\dt'}p'_{ex}
=(\ddt'+u'_{ex}\ddx')p'_{ex}=-e E'_x \; .
\label{eq:momentum} 
\eea
This same equation can be written in $S$ where, in order to obtain harmonic and non-relativistic solutions, $p_{ex} \simeq m_eu_{ex}$ and the total time derivative is linearized as ${\dd}/{\dt} \simeq \ddt$. To obtain harmonic solutions in $S'$ (where ${\dd}/{\dt'}=u'_{ex}\ddx'$), we first use the identity $\dd(\gamma(u)u)=\gamma^3(u)\dd u$ to rewrite Eq.(\ref{eq:momentum}) as
\bea
m_e\gamma^3(u'_{ex})\ddx'u'_{ex}=-e E'_x \; ,
\eea
and we write $u'_{ex}=-v_p+v'_{ex}$ with $v'_{ex}\ll c$, which yields $\gamma(u'_{ex})\simeq \gamma_p$ and $u'_{ex}\ddx'u'_{ex} \simeq -v_p\ddx'v'_{ex}$. We thus obtain
\bea
v'_{ex} \simeq -\frac{ieE_0}{m_ev_p\gamma_p^3}\E^{ik'x'} \; .
\eea
For the electron density, using $\rho'=e(\gamma_pn_0-n'_e)$ we obtain from Eq.(\ref{eq:rho-Jx})
\bea
n'_e=\gamma_pn_0-\frac{ik'E_0}{4\pi e}\E^{ik'x'} \equiv \gamma_pn_0+\delta n'_e \; .
\eea 
The current density can thus be written as 
\bea
J'_x&=&e(\gamma_pn_0v_p-n'_e(v'_x-v_p)) \nonumber \\
   &=&-e(\gamma_pn_0v'_x-\delta n'_ev_p+\delta n'_ev'_x) \; .
\label{eq:Jx2}
\eea
Since we are considering a linear wave, i.e. a linearized solution of the equations, it is consistent to neglect the nonlinear term $\delta n'_ev'_x$ in Eq.(\ref{eq:Jx2}). The condition $J'_x=0$ thus yields
\bea
\frac{4\pi e^2 n_0}{m_e\gamma_p^2v_pk'}-k'v_p=0 
\eea
i.e.  ${\omega_p^2}/{\gamma_p^2}=k^{'2}v_p^2$ which is equivalent to $v_p=\omega_p/k$. 

The frame $S'$ is also convenient to estimate the maximum energy gain for a test electron ``trapped'' in the field of the wave, as was made by Tajima \& Dawson\cite{tajimaPRL79}. In fact, in $S'$ the electron moves in the (periodic) wells of the potential energy $-e\Phi'$ where the electric potential is
$\Phi'=\Phi'(x')=(iE_0/k')\E^{ik'x'}$ since $E'_x=-\ddx'\Phi'$. The maximum energy gain is simply $W'=2eE_0/k'$, i.e. the difference between the maximum and minimum of the potential energy. The corresponding value in the laboratory can thus be obtained by transforming back the energy-momentum of the trapped electron in the $S$ (lab) frame.\footnote{It is worth noticing that this reasoning only estimates the maximum possible gain and does not take into account the issues of ``injection'' (placing the electron with zero velocity on the crest of the wave) and ``extraction'' (removing the electron at the bottom of the wave, before it loses energy by going uphill the potential).}

By using, as in Ref.\onlinecite{tajimaPRL79}, Eq.(\ref{eq:WB-NR}) for $E_0$ while assuming $v_p \simeq c$, one obtains $W' \simeq 2\gamma_pm_ec^2 \gg 1$ so that the electron momentum $p'_{x} \simeq W'/c$. The Lorentz transformation of the energy-momentum four-vector $(p'_0,p'_x) \simeq (W'/c,W'/c)$ back to $S$ yields $p_0=\gamma_p(p'_0+v_pp'_x) \simeq 2\gamma_pW'=4\gamma_p^2m_ec^2$. As we will see in section~\ref{sec:nonlinear} below, this estimate for the electron energy is greatly increased when accounting for the nonlinear character of the relativistic plasma wave.   

\section{Nonlinear case}
\label{sec:nonlinear}

The analysis of the preceding Section~\ref{sec:linear} shows that in the co-moving frame $S'$ the plasma wave is static and the total current vanishes (this latter property, albeit ultimately trivial, is probably less familiar to plasma physics experts). These two properties are not specific of the linear harmonic wave but hold in general for an electrostatic, sub-luminal wave, thus we take them as the starting condition in the search for an exact nonlinear, relativistic plane wave solution in $S'$.
We thus assume that the wave in $S'$ is described by an electrostatic potential $\Phi'=\Phi'(x')$ of arbitrary amplitude, and that the total electric current (including that due to the drifting ions) is zero. We thus have a system composed by Eqs.(\ref{eq:poisson}-\ref{eq:momentum}), written in 1D using the potential $\Phi'$, and Eq.(\ref{eq:current}):
\bea
\ddx^{'2}\Phi'&=&4\pi e(n'_e-n'_0) \; , \label{eq:dim1}\\
u'_{ex}\ddx'p'_{ex}&=&e\ddx\Phi' \; , \label{eq:dim2}\\
n'_eu'_{ex} &=& -n_0\gamma_pv_p \label{eq:dim3}
\; . 
\eea
The second equation can be easily rewritten in the form $m_ec^2\gamma(u'_x)-\Phi'=\mbox{cost.}$ which expresses energy conservation for the electron flow. Notice that, since the equations are nonlinear, all variables are taken to be real.

Before proceeding, it is useful to put the equations in a dimensionless form by measuring the distance in units of $c/\omega_p$, the density in units of $n_0$, the momenta in units of $m_ec$ and, consistently, the potential in units of $m_ec^2/e$. We thus define dimensionless variables (dropping primes for simplicity) as  
\bea
\tau\equiv\frac{\omega_px}{c} \; , \quad
N_e\equiv\frac{n'_e}{n_0} \; , \quad p\equiv\frac{p'_x}{m_ec} \; , \quad \phi\equiv\frac{e\Phi'}{m_ec^2} \; .
\eea
Eqs.(\ref{eq:dim1}), (\ref{eq:dim2}) and (\ref{eq:dim3}) thus become
\bea
\ddtau^2\phi &=&N_e-\gamma_p \ , \label{eq:poisson'}\\ 
\gamma-\phi &=& \mbox{cost.} \label{eq:energy'} \; , \\
N_e\frac{p}{\gamma}&=&-\gamma_p\beta_p \; .
\eea

\begin{figure}
\includegraphics[width=0.5\textwidth]{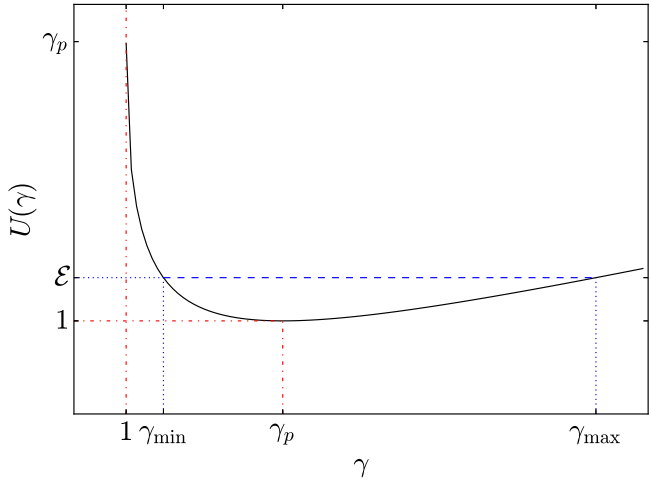}
\caption{The pseudopotential $U(\gamma)$, Eq.(\ref{eq:pseudoP}).}
\label{fig:pseudoP}
\end{figure}

By using $p=-(\gamma^2-1)^{1/2}$ (notice that $p<0$) it is straightforward to obtain for $\gamma$
\bea
\ddtau^2\gamma=\gamma_p\beta_p\frac{\gamma}{(\gamma^2-1)^{1/2}}-\gamma_p
\equiv -\frac{\partial}{\partial\gamma}U(\gamma) \; ,
\label{eq:newton}
\eea
where 
\bea
U(\gamma)=\gamma_p\left(\gamma-\beta_p(\gamma^2-1)^{1/2}\right) \; .
\label{eq:pseudoP}
\eea
Eq.(\ref{eq:newton}) has the same form as Newton's equation for a particle having mass equal to one and potential energy $U(\gamma)$. 
We can thus study the profile of $\gamma=\gamma(\tau)$ by the analogy with the motion of a pseudoparticle, with $\gamma$ playing the role of the coordinate and $\tau$ that of time, moving into the pseudopotential $U(\gamma)$ shown in Fig.\ref{fig:pseudoP}. The pseudovelocity $-\ddtau\gamma$ equals the electric field $E=-\ddtau\phi=-\ddtau\gamma$ because of Eq.(\ref{eq:energy'}).

The ``conservation of energy'' for the pseudoparticle is given by the equation
\bea
\frac{1}{2}(\ddtau\gamma)^2+U(\gamma)={\cal E} \; ,
\label{eq:pseudoE}
\eea
where the constant ${\cal E}$ plays the role of the total energy. As shown in Fig.\ref{fig:pseudoP}, if ${\cal E}<\mbox{max}[U(\gamma)]=\gamma_p$, the pseudoparticle bounces back and forth between the two points $\gamma_{\rm min}$ and $\gamma_{\rm max}$ which are the solutions of the $U(\gamma)={\cal E}$ equation. The pseudoparticle motion is thus periodical, which corresponds to a periodical waveform. If ${\cal E}>\gamma_p$, no periodical solution are found. Thus,  The limit ${\cal E}=\gamma_p$ corresponds to the maximum possible amplitude for the plasma wave, and to the maximum possible value for the pseudovelocity, i.e. the electric field $E$. This value is reached by a pseudoparticle falling from top of $U(\gamma)$ down to the bottom where $\gamma=\gamma_p$ and $U(\gamma_p)=1$, and is thus given by
\bea
E_{\rm max}=\sqrt{2}(U(1)-U(\gamma_p))^{1/2}=\sqrt{2}(\gamma_p-1)^{1/2} \; .
\label{eq:Emax}
\eea
Notice that $E_{\rm max}$ has the same value in both $S$ and $S'$ frames.
We thus obtain, putting back dimensional units, that a plasma wave cannot exceed the \emph{wave-breaking} field
\bea
E_{\rm WB}=\frac{m_ec\omega_p}{e}E_{\rm max}=\frac{\sqrt{2}m_ec\omega_p}{e}(\gamma_p-1)^{1/2} \; , \label{eq:WBfield}
\eea
as first obtained by Akhiezer \& Polovin\cite{akhiezer-polovinJETP56} by a calculation in the laboratory frame. 

\begin{figure}
\includegraphics[width=0.5\textwidth]{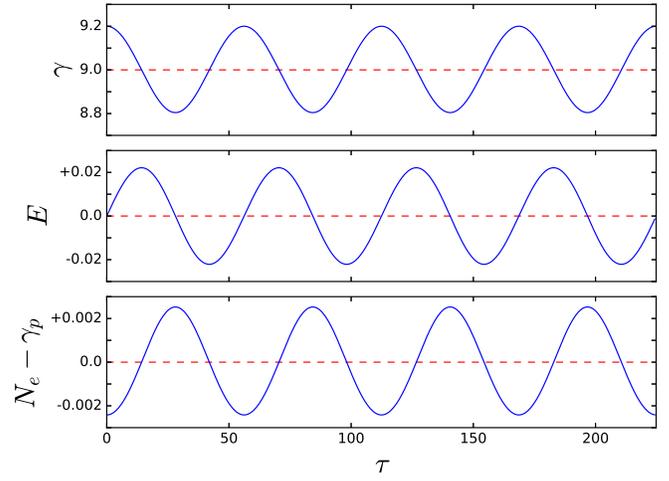}
\caption{Numerical solution for the profiles of $\gamma$, $E$ and $N_e-\gamma_p$ in the boosted frame for $\gamma_p=9$ and ${\cal E}=1.0+2.445 \times 10^{-4}$.}
\label{fig:gamma9}
\end{figure}

The formulation in terms of a pseudopotential which is a function of $\gamma$ has the advantage to immediately provide information about the energy gained by a ``test'' electron injected in the wave field, whose maximum $\gamma$-factor is simply the maximum amplitude $\gamma_{\rm max}$.  At the breaking threshold $U(\gamma_{\rm max})=\gamma_p$ (see Fig.\ref{fig:pseudoP}) and for $\gamma \gg\gamma_p$ we have $U(\gamma) \simeq \gamma\gamma_p(1-\beta_p)$. Posing $U(\gamma)=\gamma_p$ yields $\gamma=1=\gamma_{\rm min}$ or $\gamma \simeq (1-\beta_p)^{-1} \simeq 2\gamma_p^2=\gamma_{\rm max}$. To obtain the corresponding value in the laboratory frame $S$, we use the Lorentz transformation for the energy-momentum four-vector. This yields $p_0=\gamma_p(p_0'+\beta_pp'_xc)$ with $p_0'=m_ec^2\gamma_0 \simeq 2\gamma_p^2m_ec^2$, $p'_xc \simeq {\cal E}'$, and $\beta_p \simeq 1$. We thus obtain
\bea
{p}_{0{\rm max}} \simeq m_ec^2\gamma_p(2\gamma_p^2+2\gamma_p^2)=4m_ec^2\gamma_p^3 \; ,
\eea
in agreement with the result of Ref.\onlinecite{esarey-pilloffPoP95}.
Hence the estimate of Ref.\onlinecite{tajimaPRL79} for the maximum achievable energy is increased by a substantial factor $\gamma_p$.

Using the mechanical analogy, it is quite easy to obtain other properties of the plasma wave. Moreover, Eq.(\ref{eq:newton}) can be straightoforward integrated numerically, so that a student reader may compare the analytical estimates with the exact numerical solution. To start such exercise, we first check that we recover the linear, harmonic wave. The periodical waveform is sinusoidal only for small amplitude oscillations around the minimum of $U(\gamma)$ at $\gamma=\gamma_p$, such that we can expand the potential up to second order in Taylor's series:
\bea
U(\gamma) &\simeq& U(\gamma_p)+\frac{1}{2}\left.\partial^2_{\gamma}U\right|_{\gamma=\gamma_p}(\gamma-\gamma_p)^2 \nonumber \\
 &=& 1+\frac{1}{2}(\beta_p\gamma_p)^{-2}(\gamma-\gamma_p)^2 \; .
\eea 
Thus, the small amplitude oscillations have a frequency $\Omega\equiv (\beta_p\gamma_p)^{-1}$, which corresponds to a plasma wavevector $k'=(\omega_p/c)\Omega=(\omega_p/v_p)/\gamma_p=k/\gamma_p$ as obtained in Section~\ref{sec:linear}. Obviously, also the electric field $E=-\ddtau\phi=-\ddtau\gamma$ and the electron density $N_e=\gamma_p+\ddtau^2\gamma$ have a sinusoidal profile. This is confirmed by the numerical solution of Eq.\ref{eq:newton} for $\gamma_p=9$ and ${\cal E}=1.0+2.445 \times 10^{-4}$, which corresponds to an harmonic oscillation of $\gamma(\tau)$ between $\gamma_{\rm min}=8.8$ and $\gamma_{\rm max}=9.2$ with a wavelength equal to $2\pi\beta_p\gamma_p=56.2$ as shown in Fig.\ref{fig:gamma9}.

\begin{figure}
\includegraphics[width=0.5\textwidth]{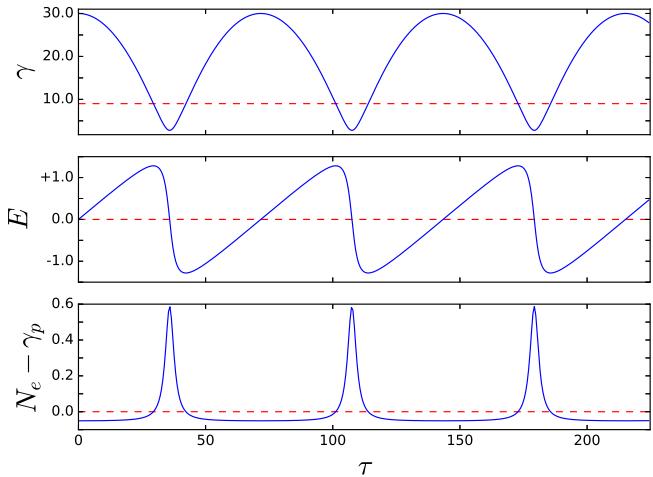}
\caption{Numerical solution for the profiles of $\gamma$, $E$ and $N_e-\gamma_p$ in the boosted frame for $\gamma_p=9$ and ${\cal E}=1.8$.}
\label{fig:gamma30}
\end{figure}

As the value of ${\cal E}$ is increased, the oscillation between $\gamma_{\rm min}$ and $\gamma_{\rm max}$ becomes anharmonic and asymmetrical. Since the pseudopotential $U(\gamma)$ gets very steep on the $\gamma<\gamma_p$ side, the pseudoparticle approaching $\gamma_{\rm min}$ undergoes a very strong pseudoacceleration at the turning point where a rapid inversion of the pseudovelocity ($E$) occurs: this corresponds to a spike in the pseudocoordinate ($\gamma$). In contrast, $U(\gamma)$ rises up gently for $\gamma>\gamma_p$, resulting in modest pseudoacceleration and the pseudoparticle spending much more ``time'' in this region.
The resulting form of $\gamma(\tau)$ has spiky minima and smooth maxima. In particular, for large values of $\gamma_p$ the pseudopotential $U(\gamma)$ is well approximated by its linear asymptotic limit $U(\gamma) \simeq \gamma_p(1-\beta_p)\gamma $ over a large part of the region accessible to the pseudoparticle motion, thus the shape of $\gamma(\tau)$ near $\gamma=\gamma_{\rm max}$ gets approximately parabolic (with a focal $\simeq 2/(\gamma_p(1-\beta_p))$), being equivalent to the motion of an heavy object starting from ground with some initial velocity. Correspondingly, the electric field $E$ acquires a ``sawtooth'' shape, i.e. it has almost a linear profile (being the derivative of a parabola) between the spikes, where it undergoes an abrupt inversion of sign. The corresponding density variation $N_e-\gamma_p$, being proportional to $-\ddtau E$, shows a periodic sequence of regions of almost constant (negative) values between sharp (positive) spikes. This qualitative analysis is confirmed by the numerical solution, shown in Fig.\ref{fig:gamma30} for $\gamma_p=9$ (as in Fig.\ref{fig:gamma9}) and ${\cal E}=1.8$, for which $\gamma_{\rm min}=2.78$ and $\gamma_{\rm max}=30.0$. The solution also shows the nonlinear increase of the wavelength with respect to the case of small amplitude, harmonic oscillations.

\begin{figure}
\includegraphics[width=0.5\textwidth]{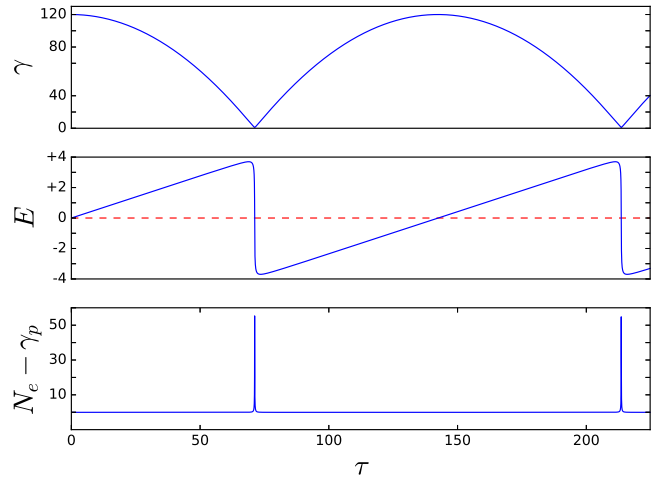}
\caption{Numerical solution for the profiles of $\gamma$, $E$ and $N_e-\gamma_p$ in the boosted frame for $\gamma_p=9$ and ${\cal E}=7.8$.}
\label{fig:gamma140}
\end{figure}

When approaching the wave-breaking limit (${\cal E}\to\gamma_p$), the pseudoparticle undergoes infinite acceleration at the cusp point $\gamma=1$, which corresponds to the density becoming singular, i.e. infinite. Fig.\ref{fig:gamma140} shows that already for ${\cal E}=7.8$ (keeping $\gamma_p=9$ as in the preceding figures) the density is extremely spiky and the electric field reaches a maximum value of $\simeq 3.7$ quite close to the prediction $E_{\rm max}=4$ for $\gamma_p=9$ (it is worth noticing that a greatly increased numerical precision is needed in this case, with respect to previous ones, in order to resolve the spikes in the density and the abrupt jumps in the electric field). 

The density $N_e(\tau)=\gamma_p+\ddtau^2\gamma(\tau)$ becomes spiky in correspondence with the peaks of $\gamma$, and it is singular at wave-breaking. For $\gamma_p\gg 1$ we have $\ddtau^2\gamma \simeq \gamma_p(1-\beta_p) \simeq 1/(2\gamma_p)$ in the parabolic valleys of the profile, thus $N_e$ has a constant value $\simeq \gamma_p-1/(2\gamma_p) \simeq \gamma_p$ for a wide region between the peaks. To obtain the corresponding value in the laboratory frame, we write the local four-current of electrons in the $\gamma_p\gg 1$ limit as
\bea
J_e^{'\mu}\simeq (-en_0\gamma_pc, -en_0\gamma_p(-c)) \; ,
\eea
since the electrons have maximum energy $\gg m_ec^2$ when crossing the valley region, and their velocity is negative.  
The back-transformation in the lab frame yields
\bea
J_e^0 &=&\gamma_p(J_e^{'0}+\beta_pJ_e^{'1}) 
\simeq \gamma_p^2(1-\beta_p)(-en_0c) \nonumber \\
&\simeq& -\frac{en_0c}{2} \; .
\eea
We obtain that at the wave-breaking limit the minimum value of the density in the valleys is $J_e^0/(-ec)=n_0/2$, which is another known feature of the exact solution\cite{akhiezer-polovinJETP56}. Notice that, with respect to the assumption of a complete density depletion which is made to estimate the non-relativistic wave-breaking threshold (\ref{eq:WB-NR}), actually the electron density does not vanish locally.

The nonlinear wavelength, i.e. the spatial period of the nonlinear wave at the wave-breaking limit can also be obtained from the pseudopotential analysis. In fact, the time spent by the pseudoparticle in the steep region $\gamma<\gamma_p$ is much shorter than that spent in the smooth region $\gamma>\gamma_p$. In this latter, since $U(\gamma)$ is fairly approximated by a linear function, the pseudoparticle moves under the action of an almost constant force $F \simeq -\gamma_p(1-\beta_p)$. Thus the pseudoparticle entering the smooth region $\gamma=\gamma_p$ at the instant $\tau=0$ with velocity $E(0)$ will return in the same point with velocity $-E(0)$ at a time $\tau_r \simeq 2E(0)/|F|$ which almost corresponds to the oscillation period if the time spent in the $\gamma<\gamma_p$ region is negligible.  
At the wave-breaking limit, we pose $E(0)=E_{\rm max}$ and use Eq.(\ref{eq:Emax}) to obtain 
\bea
\tau_r=\frac{2E_{\rm max}}{|F|}=\frac{2\sqrt{2}(\gamma_p-1)^{1/2}}{\gamma_p(1-\beta_p)} \simeq 4\sqrt{2}\gamma_p^{3/2}  \; ,
\eea
where the last approximate equality holds for $\gamma_p\gg 1$. The nonlinear wavelength at the wave-breaking limit in the $S'$ frame is thus $\lambda'_{\rm NL} \simeq (c/\omega_p)\tau_r$, while in the lab frame $S$ 
\bea
\lambda_{\rm NL}=\frac{\lambda'_{\rm NL}}{\gamma_p} \simeq 4\sqrt{2}\gamma_p^{1/2}\frac{c}{\omega_p}  \; ,
\label{eq:lambdaNL}
\eea
since the transformation factor must be the reciprocal of that for the wavevector in Eq.(\ref{eq:k'}). 
This result can be obtained in the $\gamma_p\gg 1$ limit of an exact solution (see Ref.\onlinecite{akhiezer-polovinJETP56}, Eq.(25), or Ref.\onlinecite{esareyRMP09}, also Eq.(25)) and also be checked as follows.
In $S$, the amount of electron charge (per unit surface) depleted between two peaks is $\sigma \simeq -e(n_0/2)\lambda_{\rm NL}$ at the wave-breaking limit. Such charge is concentrated into the density peaks, so that the electric field on the sides of the peak is $E_p \simeq \pm 2\pi\sigma$ as for a charged sheet with surface density $\sigma$. Posing $|E_p|=E_{\rm WB}$ and solving for $\lambda_{\rm NL}$ we obtain Eq.(\ref{eq:lambdaNL}) again.

\section{Discussion and conclusions}

In summary, we have shown that in the study of a nonlinear, relativistic cold plasma wave with sub-luminal phase velocity the equations are greatly simplified if the calculation is performed in the frame co-moving with the phase velocity. In this way, it is possible to infer the nonlinear wave profile for arbitrary amplitude and to derive important quantities without the need for higher mathematical methods. This approach may be useful for a tutorial introduction to plasma-based electron acceleration. More in general, the analysis we presented can help students to become more familiar with important topics such as the pseudopotential approach to nonlinear differential equations and the Lorentz transformations of electromagnetic fields and sources. Students interested in learning the basics of numerical computation and plasma simulation may also find interesting to reproduce the analytic predictions by a numerical integration of the pseudopotential equation (\ref{eq:pseudoP}), as we did in Figs.\ref{fig:gamma9}--\ref{fig:gamma140}, and also by the implementation of a simple plasma simulation model, such as that originally proposed by Dawson\cite{dawsonPF62} of the similar one by Boozer\cite{boozerAJP10}. For example, Fig.3.3 in our own book\cite{macchi-book} shows linear and nonlinear plasma wakefields obtained by Dawson's model.


%

\end{document}